\documentclass[useAMS,usenatbib]{mn2e}
\usepackage{graphicx}
\usepackage{multirow}
\usepackage[english]{babel}
\usepackage{caption}

\title[Collisions in young triple systems]{Collisions in young triple systems}

\author[K. Rawiraswattana, O. Lomax and S. P. Goodwin]{
    Krisada Rawiraswattana$^{1,2}$\thanks{E-mail: k.rawiraswattana@shef.ac.uk},
    Oliver Lomax$^{3}$ and
    Simon P. Goodwin$^{1}$\\
    $^{1}$Department of Physics and Astronomy, University of Sheffield, Sheffield S3 7RH\\
    $^{2}$Department of Physics, Prince of Songkla University, Hatyai, Songkla 90112, Thailand\\
    $^{3}$School of Physics and Astronomy, Cardiff University, Cardiff CF24 3AA
}

\begin{document}

\date{}
\pagerange{\pageref{firstpage}--\pageref{lastpage}} \pubyear{0000}
\maketitle
\label{firstpage}

\begin{abstract}

We perform $N$-body simulations of young triple systems consisting of two low-mass companions orbiting around a significantly more massive primary.
We find that, when the orbits of the companions are coplanar and not too widely separated, the chance of a collision between the two companions can be as high as 20 per cent.
Collisions between one of the companions (always the less massive) and the primary can also occur in systems with unequal-mass companions.
The chance of collisions is a few per cent in systems with more realistic initial conditions, such as with slightly non-coplanar orbits and unequal-mass companions.  
If the companions start widely separated then collision are very rare except in some cases when the total mass of the companions is large.  
We suggest that collisions between members of young multiple systems may explain some unusual young multiple systems such as apparently non-coeval companions.

\end{abstract}


\section{Introduction}

It appears that many, if not most, stars form in multiple systems \citep[see][]{Mathieu:1994,Goodwin:etal:2007,Goodwin:2010}.
Indeed, an increasing proportion of both pre-main sequence (PMS) and main sequence (MS) multiple systems are being found in triple or higher-order systems \citep[e.g.][]{Correia:etal:2006,Tokovinin:etal:2006,Eggleton:Tokovinin:2008,Law:etal:2010}.
There is also evidence for dynamical decay in young stellar systems suggesting that the initial triple and higher-order fraction may be even higher than the field \citep{Haisch:etal:2004,Connelley:etal:2008a,Connelley:etal:2008b}.

Observations have found a number of young stars with unusual companions, that is companions with unexpected colours \citep[][]{Hartigan:etal:1994,Koresko:etal:1997,Duchene:etal:2003,Hartigan:Kenyon:2003,Prato:etal:2003,Kraus:Hillenbrand:2009}.
The interpretation of these unusual companions is complex with some authors claiming some non-coevality between components in multiple systems \citep[e.g.][]{Hartigan:etal:1994,Hartigan:Kenyon:2003,Prato:etal:2003,Kraus:Hillenbrand:2009}, although at least some such objects may be the result of an enshrouded companion as appears to be the case in T Tauri itself \citep[see][]{Dyck:etal:1982,Ratzka:etal:2009}.  
Non-coevality is suggested by the two objects having positions on PMS tracks that cannot be explained by different masses.  
However, \citet{Baraffe:etal:2009} has suggested that apparent non-coevality may be due to differences in the accretion history of the protostars leading to different evolution along PMS tracks.  
Within a multiple system this may be due to different accretion histories (through the different angular momentum of infalling material?), but we suggest that it is possible that it could be due to a collision between low-mass companions early in the history of the system.  
Indeed, a collision has been suggested to explain the underluminosity of the low-mass brown dwarf 2M1207B by \citet{Mamajek:Meyer:2007} but described as `improbable', we will show that such a solution might not be as improbable as one might first think.

We perform $N$-body simulations of a large ensemble of triple systems to investigate the frequencies of collisions in such systems.
We describe the details of our simulations in Sec.~\ref{SECT:SIMULATIONS}.
The results are presented in Sec.~\ref{SECT:RESULTS} and followed by discussion in Sec.~\ref{SECT:DISCUSSION}.

\section{Simulations}\label{SECT:SIMULATIONS}
\subsection{Initial conditions}

We set up a three-body system of young stars in which two low-mass companions of masses $M_{2}$ and $M_{3}$ are orbiting around a primary star of (a higher) mass $M_{1}$.  
The companions are assumed to form within a circumstellar disc \citep[e.g.][]{Stamatellos:Whitworth:2009}.  
We simulate the evolution of the system once the majority of the disc has disappeared (by accretion or removed by feedback?) and so we can describe the evolution as a simple ${N}\,{=}\,{3}$ $N$-body problem.  
It is quite possible that the companions interact or collide with each other whilst a massive disc is present, and they may well interact with the disc in such a way that one or both migrate inwards or outwards.  
Such situations are extremely interesting but require modelling the hydrodynamics of the disc.

The system can be characterised by three parameters:

\subsubsection{Protostellar masses}

We simulate typical T Tauri systems with primary masses ${M_{1}}\,{=}\,{1}$ or ${2}{M_{\odot}}$.
The total masses of the companions, $M_{2}$ and $M_{3}$, have a range ${M_{2}+M_{3}}\,{=}\,{0.1-0.6}{M_{\odot}}$ in steps of ${0.05}{M_{\odot}}$, and mass ratios of ${M_{2}/M_{3}}\,{=}\,{0.25, 0.5}$ and $1$.
This gives us $33$ combinations of companion masses for each primary mass.

\subsubsection{Protostellar radii}

The radius of a young star in the systems is estimated and scaled from the mass-radius relation of low-mass main-sequence stars.
In this work, we use the empirical mass-radius relation from \citet{Caillault:Patterson:1990} which is for stars of mass ${\sim}\,{0.1-0.5}{M_{\odot}}$ in the Solar neighbourhood.
We assume that the relation extends to very low-mass stars of a few tens of Jupiter mass.
Since the PMS stars are young, their radius could be larger by a factor of ${\alpha}\,{>}\,{1}$.
The radius $R$ of a PMS star of mass $M$ can then be written as
\begin{equation}
    R = 0.92\alpha\left[\frac{M}{M_{\odot}}\right]^{0.80}R_{\odot}.
\end{equation}
The values of $\alpha$ used in the simulations are selected from $1$ up to $20$: in steps of $0.5$ for ${\alpha}\,{=}\,{1-10}$, and in steps of $1$ for ${\alpha}\,{=}\,{10-20}$.

\subsubsection{Initial positions}

Observations of PMS binaries show that they have a range of separations from sub-${\rm AU}$ to a few hundreds of ${\rm AU}$ \citep[e.g.][]{Mathieu:1994,Patience:etal:2002,Konopacky:etal:2007,Goodwin:2010}.
The peak in separations of PMS binaries appears to be around ${100-200}\,{\rm AU}$ with an excess above the field \citep[see e.g.][]{Patience:etal:2002,Konopacky:etal:2007}.  
To cover the bulk of the separation range of PMS companions we perform simulations with companions between ${20-100}\,{\rm AU}$, ${100-200}\,{\rm AU}$, and ${200-300}\,{\rm AU}$.

The two companions are placed randomly in the given ranges measured from the primary and usually both in the same plane (we also perform simulations to test the effect of non-coplanar motion by giving the third companion a small velocity component in $z$-direction).
The angular separation between the companions is forced to be greater than ${45^{\circ}}$ in order to avoid initial states which result in a collision or ejection almost immediately.
They are given the correct velocity for a circular orbit, and both orbit in the same direction.

For each value of $M_{1}$, $M_{2}$, $M_{3}$, and $\alpha$ we perform an ensemble of ${5}\,{\times}\,{10^{3}}$ or $10^{4}$ simulations.

\subsection{Numerical method}

The integrator that we use is a variable stepsize forth-order Adams-Bashforth-Moulton predictor-corrector \citep[e.g.][]{Binney:Tremaine:2008,Mathews:Fink:2004}.
This integrator requires only two force evaluations per timestep to provide a solution with high accuracy.
The fractional energy error of the calculations is kept below $10^{-5}$, and a system with unacceptable error is reintegrated from the beginning with a higher accuracy.
If an error persists the result is omitted from analyses (reducing the number of the systems in an ensemble).

\subsection{Termination criteria}

Simulations can result in either a collision, an ejection, or a stable system.

\subsubsection{Ejections}

We consider a companion `ejected' if it travels further than ${5000}\,{\rm AU}$ from the primary.
Whilst this might not always be unbound, we consider that a companion on an orbit of at least ${5000}\,{\rm AU}$ apastron distance will be very soft and likely to be perturbed by interactions with other stars in the cluster in which we assume our young multiple system has formed.

\subsubsection{Collisions}

Stars collide if the mutual separation is less than the sum of their radii as calculated above.
There are two possible types of collisions that can occur: (i) collisions between companions (companion-companion collisions, or CCCs), and (ii) collisions between a companion and the primary star (companion-primary collisions, or CPCs).

\subsubsection{Stable systems}

We consider a system stable if there are no ejections or collisions within ${5}\,{\rm Myr}$.
This length of time covers the typical ages of clusters containing T Tauri systems.
(As we shall see, stable systems are rare and so the exact length of this criterion is unimportant as long as it is ${>}\,{1}\,{\rm Myr}$).

\section{Results}\label{SECT:RESULTS}

Our fiducial system has a primary mass ${M_{1}}\,{=}\,{1}{M_{\odot}}$, and companion separations between ${100-200}\,{\rm AU}$ about the peak of the PMS separation distribution.
The main result of the fiducial system simulations is as would be expected: most of the systems decay by ejection within ${\sim}\,{100}$ crossing times, preferentially ejecting their least-massive member \citep[e.g.][]{Anosova:1986,Sterzik:Durisen:1998}.
However, we also often find a significant number of companion-companion collisions (CCCs) but very few stable systems or companion-primary collisions (CPCs).  
In ensembles with different initial conditions stable systems and CPCs can become important, however dynamical decay is most often by far the dominant outcome (see below).

In a three-body system such as the ones we are simulating there is one major body (the primary) which dominates the potential.
At sufficiently low-mass companions would barely feel each-other's gravity and would evolve as a planetary system (${M_{1}}\,{\gg}\,{M_{2}+M_{3}}$).
As the mass of the companions increases they will perturb each-other's orbits, but always in the global potential of the primary (${M_{1}}\,{>}\,{M_{2}+M_{3}}$).
As their orbits are perturbed by each other the companions will undergo a number of close encounters which will transfer energy from one body to the other.
Often an encounter will provide one of the companions with enough energy to escape the system altogether -- an ejection.
However, a not insignificant fraction of encounters will be close enough to cause a collision.
This collision is almost always between the two minor bodies in the system as they are the bodies whose orbits are most perturbed by the other.

\subsection{Companion-companion collisions}

\begin{figure}
    \centering
    \includegraphics[angle=270,width=84mm]{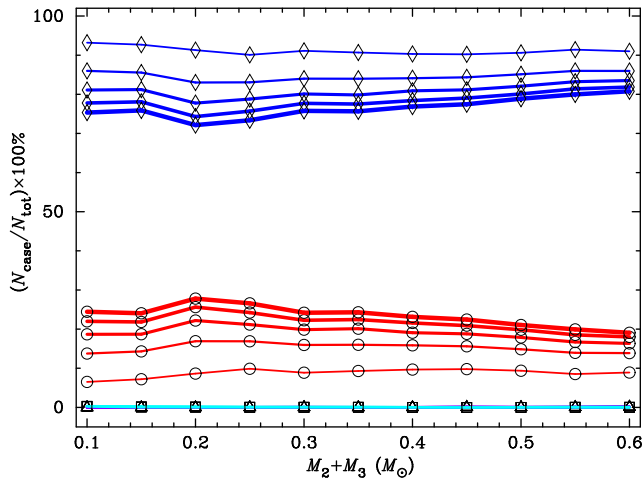}
    \caption{
        The effect of changing the radius factor (${\alpha}$) in our fiducial system (with primary mass ${1}{M_{\odot}}$ and orbits between ${100-200}\,{\rm AU}$) on the frequencies of ejections (blue lines with diamond symbols), CCCs (red lines with circle symbols), CPCs (purple lines with triangle symbols), and stable systems (cyan lines with square symbols).
        The frequencies are plotted against total companion mass $M_{2}+M_{3}$ of the ensembles.
        In all cases ${M_{2}}\,{=}\,{M_{3}}$.
        Increasing line thicknesses correspond to increasing radius factors ${\alpha}\,{=}\,{1, 5, 10, 15, 20}$.
        Note that the fractions of CPCs and stable systems are negligible.
    }
    \label{FIG:EFF-OF-ALPHA}
\end{figure}

In Fig.~\ref{FIG:EFF-OF-ALPHA} we show the change in the number of ejections (lines with diamond symbols) and collisions (lines with circle symbols) for our fiducial system with increasing total companion mass where both companions are of the same mass (${M_{2}}\,{=}\,{M_{3}}$).
The increasing thickness of the lines shows the effect of increasing the radii of all the stars by a factor ${\alpha}\,{=}\,{1, 5, 10, 15, 20}$ (moving upwards with increasing thickness to higher $\alpha$ for CCC, and downwards for ejections).

Unsurprisingly, larger stellar radii result in more collisions, from a few per cent when ${\alpha}\,{=}\,{1}$ (the MS radius), to around 20 per cent when ${\alpha}\,{=}\,{20}$ (a very large PMS star). 
Collision timescales are similar to those of ejections and usually occur within ${\sim}\,{100}$ crossing times.

The fraction of CCCs stays relatively constant for any given $\alpha$ as $M_{2}+M_{3}$ changes from $0.1$ to ${0.6}{M_{\odot}}$ (from 10 to 60 per cent of the primary mass).
This is because even at ${0.6}{M_{\odot}}$, the companions are still not the dominant contributor to the potential.
And even though more massive companions require a greater escape energy (as they are more massive and escaping from a more massive system), the energy change in an encounter between more massive companions is greater by the same order.

The number of CCCs changes in an unusual way with increasing $\alpha$.
The number of CCCs increases quite rapidly as $\alpha$ is increased from $1$ to $5$ to $10$, but the relative increase in the number of collisions as $\alpha$ is increased from $10$ to $15$ to $20$ is small.

\begin{figure}
    \centering
    \includegraphics[angle=270,width=84mm]{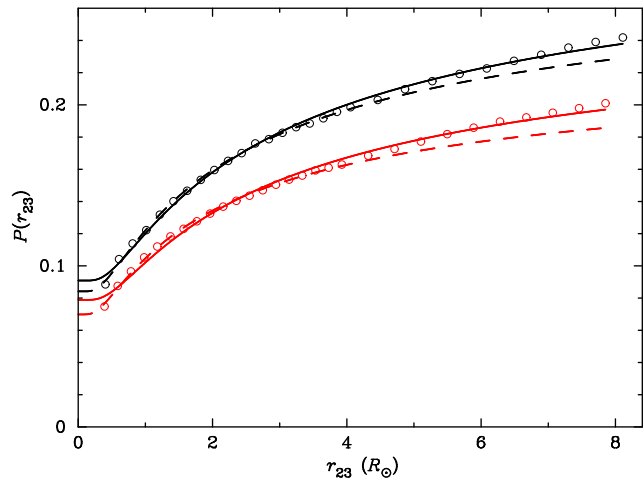}
    \caption{
        The probability of an encounter at a companion separation ${\leq}\,{r_{23}}$, $P(r_{23})$, as measured by the frequency of CCCs with increasing $\alpha$.
        In all cases ${M_{1}}\,{=}\,{1}{M_{\odot}}$ and ${M_{2}+M_{3}}\,{=}\,{0.3}{M_{\odot}}$.
        Black circles are for systems with mass ratio ${M_{2}/M_{3}}\,{=}\,{1}$ and red circles for ${M_{2}/M_{3}}\,{=}\,{0.25}$.
        Solid lines are the fits from Eqn.~(\ref{EQN:LEVY-CDF}) for all data, whilst dashed lines show the fits for only ${\alpha}\,{=}\,{1-10}$.
    }
    \label{FIG:CDF}
\end{figure}

A CCC occurs when the separation of the companions $r_{23}$ is equal to the sum of their radii ${R_{2}+R_{3}}$.
Therefore, the number of CCCs is the probability that the separations become less than $r_{23}$, $P(r_{23})$.
Fig.~\ref{FIG:CDF} shows the increase in $P(r_{23})$ with $r_{23}$ for two example systems with ${M_{2}+M_{3}}\,{=}\,{0.3}{M_{\odot}}$ and mass ratios ${M_{2}/M_{3}}\,{=}\,{0.25}$ and ${1}$.

The trends in Fig.~\ref{FIG:CDF} can be fitted very well with the L\'{e}vy cumulative distribution function (CDF) of the form
\begin{equation}\label{EQN:LEVY-CDF}
    P(r_{23}) = {a}\,{\rm erfc}{\left(\sqrt{b/r_{23}}\right)}+c,
\end{equation}
where $a$, $b$ and $c$ are constants, and ${\rm erfc}$ is the complementary error function.
This CDF provides the best fit with meaningful properties, namely, ${dP(r)/dr}\,{=}\,{0}$ at ${r}\,{=}\,{0}$ and $\lim\limits_{{r}\,{\to}\,{\infty}}\,{P(r)}\,{=}\,{\rm constant}$.
The values of the constants and coefficient of determination (${\mathcal{R}^{2}}$) corresponding to each data set are summarized in Table \ref{TAB:FIT-VALS}.
The solid lines in Fig.~\ref{FIG:CDF} show the best fits to high-$\alpha$, while the dashed lines show the best fits to low-$\alpha$ (see below).

\begin{table}
    \begin{center}
        \caption{
            The constants in Eqn.~(\ref{EQN:LEVY-CDF}) obtained from nonlinear regressions of the data in Fig.~\ref{FIG:CDF}.
            The goodness of fit is represented by the coefficient of determination ${\mathcal{R}^{2}}$ (closer to $1$ is better).
        }
        \begin{tabular}{c c c c c c}
            \hline\hline
            $\alpha$ & $M_{2}/M_{3}$ & $a$ & $b$ & $c$ & ${\mathcal{R}^{2}}$\\
            \hline
            \multirow{2}{*}{1-20}
                & 1.00 & 0.254 & 1.242 & 0.091 & 0.9971\\
                & 0.25 & 0.207 & 1.265 & 0.079 & 0.9945\\
            \hline
            \multirow{2}{*}{1-10}
                & 1.00 & 0.232 & 0.972 & 0.084 & 0.9988\\
                & 0.25 & 0.180 & 0.847 & 0.070 & 0.9984\\
            \hline
        \end{tabular}
        \label{TAB:FIT-VALS}
    \end{center}
\end{table}

That the data are well fitted by the L\'{e}vy distribution can be explained by assuming the following.
(i) As the distance between the companions at collision is far smaller than the distance to the primary the encounter between the two companions is approximately a two-body problem.
(ii) The effects of a rotating frame of reference about the primary are small at low $r_{23}$.
The problem can then be simplified to one body of a reduced mass ${\mu}\,{=}\,{M_{2}M_{3}/(M_{2}+M_{3})}$ orbiting around a fixed central mass of ${M}\,{=}\,{M_{2}+M_{3}}$.
Let us now consider $r_{23}$ as the separation between the masses $\mu$ and $M$ at the pericentre of the orbit.
At this turning point, as the radial velocity is zero, we assume further that (iii) the tangential velocity ($v_{\rm t}$) of the mass $\mu$ follows the one-dimensional Maxwell-Boltzmann distribution, i.e. ${f(v_{\rm t})}\,{\propto}\,{\exp(-v_{\rm t}^2/2\sigma^{2})}$, where $\sigma$ is the velocity dispersion.
The probability of the mass $\mu$ having $v_{\rm t}$ from that for a circular orbit to hyperbolic orbits around the fixed mass $M$ may be written as
\begin{equation}\label{EQN:CDF1}
    P(v_{\rm t} \geq v_{\rm cir}) \propto \int_{v_{\rm cir}}^{\infty} e^{-v_{\rm t}^2/2\sigma^{2}} dv_{\rm t} \propto {\rm erfc}{\left(v_{\rm cir}/\sqrt{2}\sigma\right)}.
\end{equation}
Substituting ${v_{\rm cir}}\,{=}\,{\sqrt{GM/r_{23}}}$ in Eqn.~(\ref{EQN:CDF1}), we have
\begin{equation}
    P(r_{23}) = {a}\,{\rm erfc}{\left(\sqrt{b/r_{23}}\right)},
\end{equation}
where $a$ and $b$ are constants. The constant $c$ in Eqn.~(\ref{EQN:LEVY-CDF}) is due to `head-on' collisions in which $v_{\rm t}$ is ${\sim}\,{0}$.

In addition to the solid lines in Fig.~\ref{FIG:CDF}, the dashed lines show the fits of the data with ${\alpha}\,{\leq}\,{10}$.
Although these curves are formally better fits (higher ${\mathcal{R}^{2}}$, as shown in Table \ref{TAB:FIT-VALS}) they tend to diverge from the data at higher $\alpha$.
These divergences probably indicate some complicated dynamics that we have not included in our derivation above (i.e. the effects of motion in a rotating frame of reference).

The differential of $P(r_{23})$ with respect to $r_{23}$, $p(r_{23})$, is the probability density function (PDF) of the encounter having a separation at pericentre of $r_{23}$,
\begin{equation}
    p(r_{23}) = a\sqrt{\frac{b}{\pi}}\frac{e^{-b/r_{23}}}{r_{23}^{3/2}}.
\end{equation}

\subsection{Systems of unequal-mass companions}

The companion mass ratio (${M_{2}/M_{3}}$) plays an important role in the energy redistribution during close encounters between the companions.
In $N$-body systems, the objects in the systems tend to distribute kinetic energy equally (equipartition).
The systems thus usually consist of slow-moving-high-mass and fast-moving-low-mass objects.

\begin{figure}
    \centering
    \includegraphics[angle=270,width=84mm]{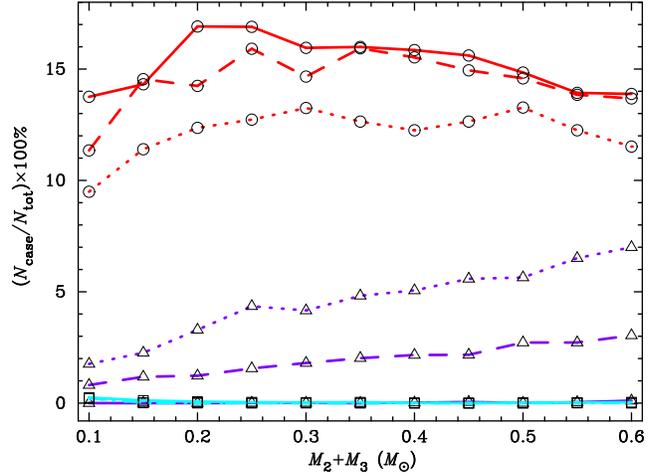}
    \caption{
        The effect of changing the companion mass ratio ($M_{2}/M_{3}$) on the frequencies of CCCs (red lines with circle symbols), CPCs (purple lines with triangle symbols), and stable systems (cyan lines with square symbols).
        Different line styles represent different mass ratios: ${M_{2}/M_{3}}\,{=}\,{0.25}$ (dotted lines), ${0.5}$ (dashed lines), and ${1.0}$ (solid lines).
        In all cases ${M_{1}}\,{=}\,{1}{M_{\odot}}$ and ${\alpha}\,{=}\,{5}$.
    }
    \label{FIG:EFF-OF-MR}
\end{figure}

In our three-body system, most of the time only the companions are closely interacting with each other (see above).
For systems with unequal-mass companions (${M_{2}/M_{3}}\,{<}\,{1}$), the equipartition of the kinetic energy usually causes the lower-mass companion to be ejected from the system.
For systems with equal-mass companions the companions need more close encounters before energy exchange is able to eject one object.
The chance of collisions in the equal-mass systems is therefore higher than in unequal-mass systems.
In Fig.~\ref{FIG:EFF-OF-MR} we can see that the number of collisions between companions (marked with circles) increases with the increasing mass ratio from ${M_{2}/M_{3}}\,{=}\,{0.25}$ (the dotted line) to ${M_{2}/M_{3}}\,{=}\,{1}$ (the solid line).

Figure.~\ref{FIG:EFF-OF-MR} also shows an interesting feature of systems with unequal-mass companions: increasing numbers of companion-primary collisions (CPCs).
We find that it is almost always the lower-mass companion that collides with the primary star.
This occurs after the lower-mass companion has been regularly perturbed by the more-massive companion orbiting the primary with a larger orbit.
Some angular momentum is extracted from the lower-mass companion during each close encounter.
This causes the orbit of the lower-mass companion to become more and more eccentric until eventually colliding with the primary star.
CPCs are almost never seen with equal-mass companions.

\subsection{The effect of other parameters}

Apart from the radius factor ($\alpha$) and the companion mass ratios, three other parameters can also affect the results.  
Firstly, the primary mass and the separation from the primary to the companions (essentially the potential energy of the system).  
Secondly, the distance between the companions.  
And thirdly, the coplanarity of the orbits.

\subsubsection{Potential energy}

\begin{figure}
    \centering
    \includegraphics[angle=270,width=84mm]{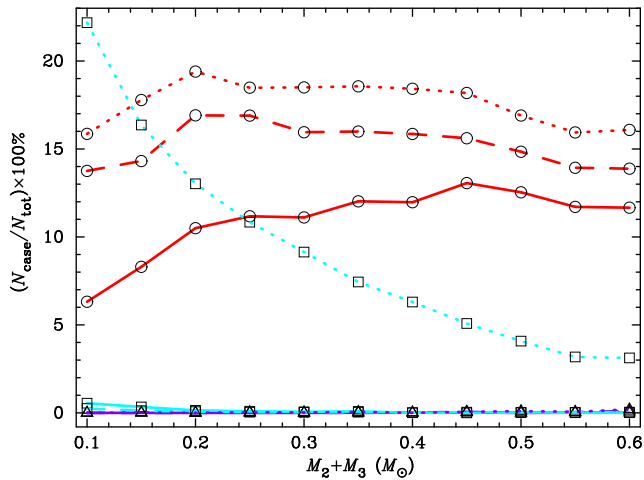}
    \caption{
        The effect of changing the companion initial orbital range on the frequencies of CCCs (red lines with circle symbols), CPCs (purple lines with triangle symbols), and stable systems (cyan lines with square symbols).
        Different line styles represent different ranges of the orbital radii: ${20-100}\,{\rm AU}$ (dotted lines), ${100-200}\,{\rm AU}$ (dashed lines), and ${200-300}\,{\rm AU}$ (solid lines).
        In all cases ${M_{1}}\,{=}\,{1}{M_{\odot}}$, ${M_{2}/M_{3}}\,{=}\,{1}$, and ${\alpha}\,{=}\,{5}$.
    }
    \label{FIG:EFF-OF-POT}
\end{figure}

The potential energy of the system is set by the primary mass and the distance of the companions from the primary.  
The deeper the potential well, the more difficult it is for an ejection to occur (the more energy needed to be transferred).  
Therefore, the deeper the potential well the more encounters are needed before an ejection occurs, and the greater the chance of a collision or the `partial' ejection of a companion to a wider stable orbit.  
In Fig.~\ref{FIG:EFF-OF-POT} we decrease the separation range of companions from ${200-300}\,{\rm AU}$ (solid lines), to ${100-200}\,{\rm AU}$ (dashed lines), to ${20-100}\,{\rm AU}$ (dotted lines).  
The number of CCCs (red lines with circles) increases with decreasing separation (increasing potential energy).  
In the case of ${20-100}\,{\rm AU}$ separations the number of stable systems is also significant (dotted cyan line with squares), indeed, at low companion masses stable systems outnumber CCCs.
Similar effects are also seen with increasing primary mass (not illustrated).

\subsubsection{Separation between the companions}

Initial separation is an important parameter.  
Frequent interactions leading to ejections and collisions occur far less often in systems in which the companions start widely separated.  
We demonstrate this with three ensembles whose companion's initial orbits are each restricted to be within two widely-separated ranges of ${r_{2}}\,{=}\,{50-60}\,{\rm AU}$ and ${r_{3}}\,{=}\,{200-210}\,{\rm AU}$: (i) ${M_{2}/M_{3}}\,{=}\,{0.5}$ (the less massive in the closer orbit), (ii) ${M_{2}/M_{3}}\,{=}\,{1}$, and (iii) ${M_{2}/M_{3}}\,{=}\,{2}$ (the more massive in the closer orbit).
  
\begin{figure}
    \centering
    \includegraphics[angle=270,width=84mm]{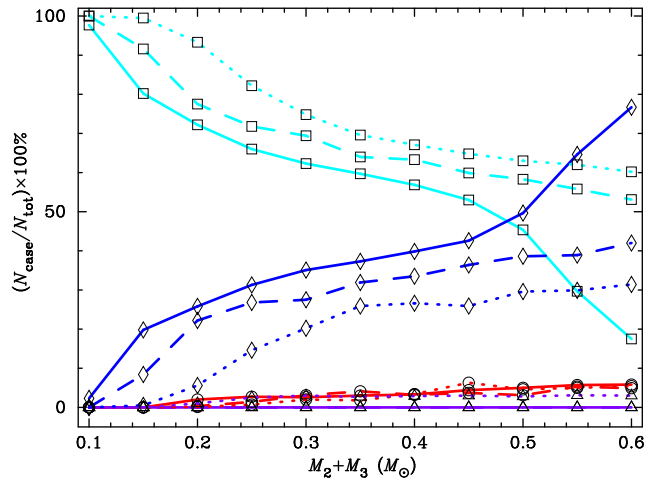}
    \caption{
        The effect of large initial separations between the companions ($r_{23}$) on the frequencies of ejections (blue lines with diamond symbols), CCCs (red lines with circle symbols), CPCs (purple lines with triangle symbols), and stable systems (cyan lines with square symbols).
        Dotted lines represent the systems with ${M_{2}/M_{3}}\,{=}\,{0.5}$ (the less massive in the smaller orbit), dashed lines with ${M_{2}/M_{3}}\,{=}\,{1}$, and solid lines with ${M_{2}/M_{3}}\,{=}\,{2}$ (the more massive in the smaller orbit).
        In all cases ${M_{1}}\,{=}\,{1}{M_{\odot}}$ and ${\alpha}\,{=}\,{5}$.
    }
    \label{FIG_FIG:EFF-OF-LARGE-r23}
\end{figure}

We can see from Fig.~\ref{FIG_FIG:EFF-OF-LARGE-r23} that most systems that start widely separated are stable.  
Systems tend to be more stable if the lower-mass companion has the closer orbit, or if ${M_{2}/M_{3}}\,{<}\,{1}$; compare the dotted cyan line (${M_{2}/M_{3}}\,{=}\,{0.5}$) with the dashed cyan line (${M_{2}/M_{3}}\,{=}\,{1}$) and the solid cyan line (${M_{2}/M_{3}}\,{=}\,{2}$) in Fig.~\ref{FIG_FIG:EFF-OF-LARGE-r23}.
Similarly to our fiducial systems, CPC apparently occurs only in the systems with ${M_{2}/M_{3}}\,{<}\,{1}$ (the dotted purple line with triangle symbols).

Although the number of CCCs drops significantly, they do not disappear entirely except for very low companion masses.  
It is unclear what the initial architecture of young triple systems would most often be.
Disc fragmentation would be expected to occur at large radii \citep[$>100$~AU,][]{Stamatellos:Whitworth:2009}, however migration within the disc may rapidly separate companions.

\subsubsection{Coplanarity of the orbits}

\begin{figure}
    \centering
    \includegraphics[angle=270,width=84mm]{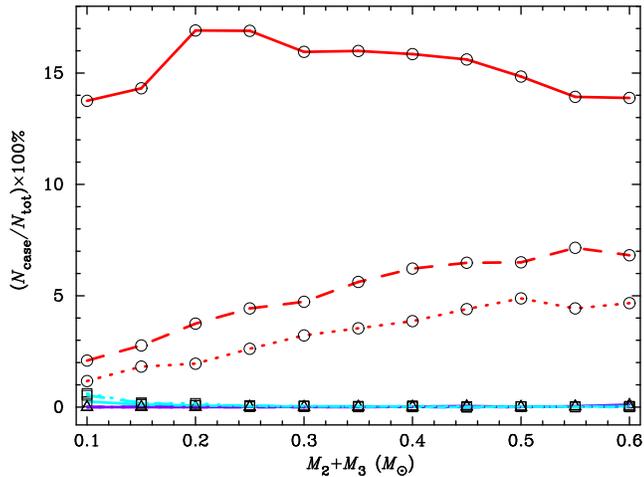}
    \caption{
        The effect of changing the orbital coplanarity on the frequencies of CCCs (red lines with circle symbols), CPCs (purple lines with triangle symbols), and stable systems (cyan lines with square symbols).
        Different line styles represent different degrees of orbital inclination between the two companions: coplanar orbits (solid lines), ${\sim}\,{5^{\circ}}$ inclined orbits (dashed lines), and ${\sim}\,{10^{\circ}}$ inclined orbits (dotted lines).
        In all cases ${M_{1}}\,{=}\,{1}{M_{\odot}}$, ${M_{2}/M_{3}}\,{=}\,{1}$, and ${\alpha}\,{=}\,{5}$.
    }
    \label{FIG_FIG:EFF-OF-NONCOP}
\end{figure}

To test the effect of non-coplanarity we perform some ensembles of simulations in which the companion $M_{3}$ is given a small velocity component in the $z$-direction just enough to make its orbit initially inclined by around (a) ${\sim}\,{5^{\circ}}$ and (b) ${\sim}\,{10^{\circ}}$.
We find that the number of collisions drops significantly, as shown in Fig.~\ref{FIG_FIG:EFF-OF-NONCOP} from  ${\sim}\,15$ per cent at zero inclination (solid line at top), to a ${\la}\,{5}$ per cent at ${5^{\circ}}$ and ${10^{\circ}}$ inclinations (dashed and dotted lines).  
Non-coplanar interactions produce fewer collisions due to the introduction of a third dimension, however there are still a non-negligible number of collisions.

\section{Discussion}\label{SECT:DISCUSSION}

We have studied the dynamical evolution of coplanar triple systems of young stars.
We find that collisions between members are not {\em very} unusual and can occur up to 20 per cent of the time in coplanar systems with two equal-mass companions.
Collisions are mostly between the two companions rather than between a companion and the primary.
We unsurprisingly find that stars with larger radii are more likely to collide, but that collisions are not uncommon at stellar radii only a few times the main sequence radius.

We find that in more realistic situations with different mass companions and slightly non-coplanar orbits the number of collisions is significantly lower, but can still be a few per cent.

We conclude that collisions in young triple systems would not usually occur, but that they might happen often enough that they could explain some unusual systems such as some of the apparently non-coeval T Tauri systems that have been observed \citep{Koresko:etal:1997,Duchene:etal:2003,Hartigan:etal:1994,Hartigan:Kenyon:2003,Prato:etal:2003,Kraus:Hillenbrand:2009}.

In our simple $N$-body simulations we are unable to determine what would happen after a collision and what the product or products might be.  
Many collisions may be glancing collisions stripping material off of one or both objects, or collisions may cause the objects to merge to make a new, larger, object \citep{Laycock:Sills:2005}.  

Whatever happens, a collision would be expected to change the mass of the final object(s) either stripping material or forming a new object.  
Such rapid and violent mass changes would be expected to change the structure of the objects and cause them to be in unexpected places for their ages on the HR diagram \citep{Baraffe:etal:2009}.  
Stripped or ejected material could persist around one or other object causing it to have a different (and unusual?) extinction to other objects in the system (an explanation for T Tauri itself? See \citealp{Ratzka:etal:2009}).  

Without detailed hydrodynamic simulations and radiative transfer modelling it is impossible to know what the colours of the new object(s) after collision might be.  
\citet{Kraus:Hillenbrand:2009} find non-coevality with the most massive star appearing younger.   
If the mass determinations are accurate then this might suggest CPCs above CCCs (despite CCCs being more common in most of our simulations).  However, the collision product will presumably be larger and so appear more massive, but it may also be hotter and so appear older?   
The collision product will presumably follow a different path towards the main sequence than it would previously have done, not appearing the `correct' colour and magnitude until it reaches the main sequence.  Without detailed modelling it is very difficult to guess how a PMS collision product would look or evolve, but it seems reasonable to assume that it would very different to a `normal' PMS star of the same (post-collision) mass.

In summary, collisions would be expected in some young triple systems.
They are not common, but are not so rare as to be ignored in attempting to explain some unusual systems \citep[e.g.][]{Mamajek:Meyer:2007}.

\section*{Acknowledgments}

Simulations in this work has been performed on Iceberg, the High Performance Computing server at the University of Sheffield.
KR acknowledges financial support from Prince of Songkla University, Thailand.
We thank the referee for their very helpful suggestions in improving the manuscript.


\end{document}